\documentclass[pre,twocolumn,floatfix,amsmath,amssymb,superscriptaddress,longbibliography]{revtex4-2}
\usepackage[utf8]{inputenc}
\usepackage{graphicx}
\usepackage{dsfont}
\usepackage{bm}
\usepackage[hidelinks]{hyperref}

\begin{document}

\title{Microscopic and collective signatures of feature learning in neural networks}

\author{A. Corti}
\affiliation{Università degli Studi di Milano, Via Celoria 16, 20133 Milano, Italy}
\affiliation{I.N.F.N. sezione di Milano, Via Celoria 16, 20133 Milano, Italy}

\author{R.~Pacelli}
\affiliation{I.N.F.N. sezione di Padova, Via Marzolo 8, 35131, Padova, Italy}

\author{P.~Rotondo}
\affiliation{Dipartimento di Scienze Matematiche, Fisiche e Informatiche,
Università degli Studi di Parma, Parco Area delle Scienze, 7/A 43124 Parma, Italy}

\author{M.~Gherardi}
\email[Correspondence email address: ]{marco.gherardi@unimi.it}
\affiliation{Università degli Studi di Milano, Via Celoria 16, 20133 Milano, Italy}
\affiliation{I.N.F.N. sezione di Milano, Via Celoria 16, 20133 Milano, Italy}

\begin{abstract}
Feature extraction --- the ability to identify relevant properties of data ---
is a key factor underlying the success of deep learning.
Yet, it has proved difficult to elucidate its nature within existing predictive theories, to the extent that there is no consensus on the very definition of feature learning.
A promising hint in this direction comes from previous phenomenological observations of quasi-universal aspects in the training dynamics of neural networks, displayed by simple properties of feature geometry.
We address this problem within
a statistical-mechanics framework for Bayesian learning in one hidden layer neural networks
with standard parameterization.
Analytical computations in the proportional limit
(when both the network width and the size of the training set are large)
can quantify fingerprints of feature learning, both collective ones
(related to manifold geometry) and microscopic ones (related to the weights).
In particular,
(i) the distance between different class manifolds in feature space is a nonmonotonic function of the temperature,
which we interpret as the equilibrium counterpart of a phenomenon observed under gradient descent (GD) dynamics, and
(ii) the microscopic learnable parameters in the network undergo a finite data-dependent displacement
with respect to the infinite-width limit, and develop correlations.
These results indicate that
nontrivial feature learning is at play in a regime where the posterior predictive distribution
is that of Gaussian process regression with a trivially rescaled prior.
\end{abstract}

\maketitle

\begin{figure*}
    \centering
    \includegraphics[width=1.\linewidth]{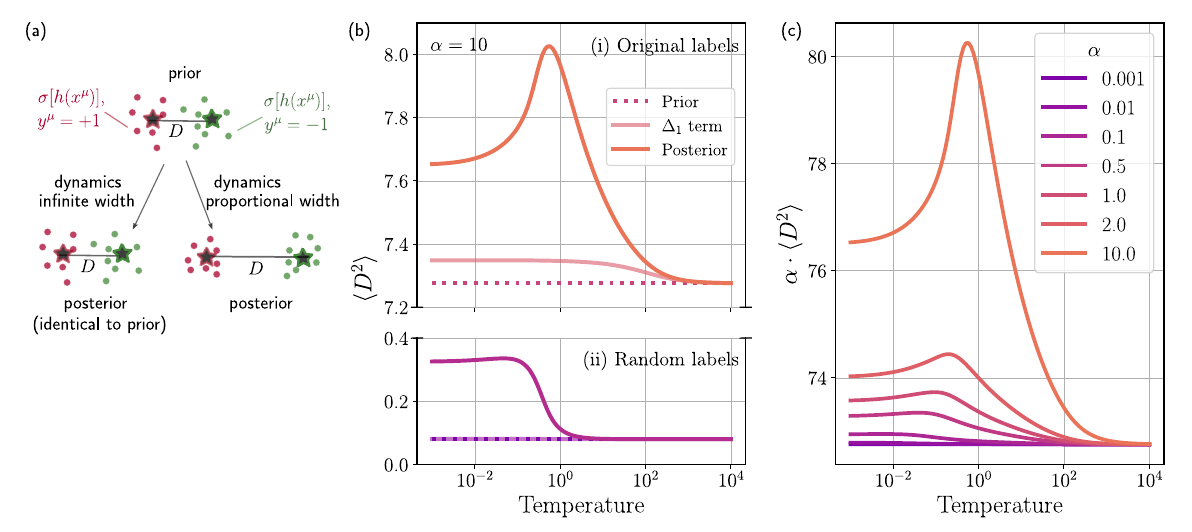}
    \caption{\textbf{(a) Separation of class manifolds in the lazy training infinite-width vs proportional regime.} In the infinite-width limit, the separation between class manifolds $D^2$ is unaffected by training. The weight displacement that occurs in this setting is not enough to produce changes in the collective observable $D^2$. In the proportional regime, $D^2$ undergoes a finite shift, due to the microscopic weight displacements.
    \textbf{(b,c) The nonmonotonicity of the distance is a signature of feature learning at proportional width.}
    The average squared distance ($y$ axis), Eq.~(\ref{eq:D2_approx}), is a nonmonotonic function of the temperature $T$ ($x$ axis). The peak is not present for randomized labels (dashed line) or in the temperature-independent prior (dotted line).
    \textbf{(c) Moving away from the infinite-width limit (larger load $\alpha$) makes the peak more prominent.} The distance ($y$ axis) is rescaled by $\alpha$ so that the infinite-width limit is finite.
    All curves are computed using the first 1000 CIFAR10 samples, split into ``airplane'' and ``automobile'' classes, with activation function $\sigma = \text{erf}$.}
    \label{fig:distance}
\end{figure*}

\emph{Introduction---}The empirical success of deep learning is fundamentally linked to the ability of neural networks (NNs) to extract meaningful features from raw data \cite{hinton86, AlexNet,krizhevsky2009learning}. Nevertheless, the mechanistic definition of such process, referred to as \textit{feature learning}, remains debated. Even for simple models, such as standard-scaled fully-connected (FC) NNs, there is no agreed-upon set of observables (properties) of their trainable parameters that definitively describe feature learning \cite{tensoriv, seroussi2023natcomm, yang2023theory,springblock, van2025coding}.

Recent progress on the theoretical analysis of NNs has established an equivalence between
standard-scaled Bayesian FC NNs and
kernel regressions,
particularly in two limits.
(i) In the infinite-width framework, where the width of the hidden layers $N_1$ is much larger than the number of training examples $P$, standard-scaled NNs are known to be mathematically equivalent to fixed Gaussian Processes (GPs), which only depend on the \textit{prior} statistics of the weights \cite{Neal,mackay1998introduction,rasmussen2006,LeeGaussian,novak2019bayesian,tensori, canatar2021,aroracnn2019, favaro2025quantitative}. In this case, training produces an infinitesimal displacement of both the network’s weights and the hidden-layer features, which is enough to fit the dataset examples. (ii) In the less overparameterized proportional regime, where $P/N_1 = \alpha >0$, one lacks such formal equivalence, yet GPs still come to hand when analysing predictive performance. Specifically, the average generalisation error of the NN is found to be the same as that of a GP regression with a set of free hyperparameters that are fine-tuned to the task at hand. Differently from the infinite-width limit, this GP is not fixed at initialisation, and depends on the \textit{posterior} statistics of the weights \cite{SompolinskyLinear, doi:10.1073/pnas.2301345120, pacelli2023statistical,pacelli2024GP,cui2023optimal, camilli2025information,aiudi2023,  JMLR:v26:24-1158, baglioni2024predictive, 9pkk-d4bm}.
(This regime allows for the analysis of continual learning and transfer learning as well
\cite{shan2024order,ingrosso2024tl}.)

These descriptions in terms of GPs suggest that looking at the predictive distribution may not be enough to characterise feature learning in FC models. In fact, consistent experimental evidence points to the fact that finite, yet overparametrized, FC networks are eventually outperformed by a suitable GP \cite{neurips_empirical_study} (at least in most computer vision tasks), prompting the identification of a different set of observables to describe feature learning  in this context.

A key insight emerges from recent empirical studies of feature dynamics in FC NNs. Experiments show a robust behaviour that occurs during training: collective observables of the features are nonmonotonic in training time, with a quasi-universal inversion point that is consequential for good generalization properties \cite{Ciceri2024, PukowskiLu2024}. More precisely, for binary classification problems, the class manifolds (the two sets containing the representations in feature space of data with the same label \cite{PhysRevX.8.031003, PhysRevLett.125.120601, PhysRevE.102.032119, Cohen2020, Pastore_2021}) initially become well separated, and then approach each other when the network is learning the most ``challenging'' data samples. These results, in addition to pre-existing numerical evidence \cite{Farrell2022,kamnitsas2018semisupervised}, indicate that the geometry of class manifolds in feature space is intimately linked to feature extraction.

In this manuscript, we investigate collective and microscopic equilibrium observables, linked to hidden-layer features, in
an analytically tractable theory of Bayesian one-hidden-layer (1HL) fully connected neural networks in the proportional regime.
We compute the posterior average squared distance $\langle D^2\rangle $ between class manifolds in a binary classification problem, and analyse the second order statistics of the hidden layer weights. The separation of internal representations across classes has been widely studied as a key mechanism underlying generalization and robustness in deep learning \cite{Xie2025LCL,springblock,Ciceri2024}. In this context, we provide an analytical characterization of representation separation in a Bayesian setting.
Our results can be summarised in three points.
(i) The posterior distance remains unaltered in the infinite-width limit, while it departs from its prior value at proportional width (see sketch in panel (a) of Fig.~\ref{fig:distance}).
(ii) The observable $\langle D^2\rangle $ is a non-monotonic function of the Gibbs temperature $T$. The nonmonotonicity disappears in the infinite-width limit, or when information about the features is removed from the data by randomising the labels.
(iii) The hidden-layer weights become correlated in the proportional regime, undergoing a finite displacement, differently from the infinite-width limit, where the parameter statistics is known to be unaffected by training.

\emph{Setting of the learning problem---} We consider
a training set
$\mathcal{D} = \{x^\mu,y^\mu\}_{\mu=1}^P$, with $x^\mu\in \mathbb R^{N_0}$ and $y^\mu\in\mathbb R$, and a 1HL FC network
$f(x) = v \cdot \sigma [h(x)]/\sqrt{N_1}$, where the pre-activations read $h(x) = wx/\sqrt{N_0}$, and $\sigma$ is a pointwise nonlinear activation function.
The matrix $w \in \mathbb R^{N_1 \times N_0}$ and vector $v \in \mathbb R^{N_1}$ contain the microscopic degrees of freedom of the system. We refer to them collectively as $\theta=\{v,w\}$.
Multiplying the network's output by $1/\sqrt{N_1}$ corresponds to the standard scaling \cite{tensoriv}. For the sake of simplicity, we restrict our analysis to zero-mean activation functions (e.g., erf, tanh).
The statistics of $v$ and $w$ in the Bayesian setting are determined by their priors, which we take to be rescaled normals,
and by a likelihood function, for which we use the mean-squared error loss.
The properties of the Bayesian network are then determined by the partition function
$ \mathcal{Z}(X,y)  = \int \text{d}\mu(\theta)\, \text{exp}(-\beta||f_\theta(X)-y||^2/2)$.
The shorthand notation $f_\theta(X) = (f_\theta(x^\mu), x^\mu \in \mathcal{D})$ indicates the collection of the network's outputs to the training samples, while
the measure d$\mu(\theta)$ indicates
integration over the prior on the weights $w\sim\mathcal{N}(\bm{0},\mathds{1}/\lambda_0)$, $v\sim\mathcal{N}(\bm{0},\mathds{1}/\lambda_1)$.
As shown in \cite{SompolinskyLinear, doi:10.1073/pnas.2301345120,JMLR:v26:24-1158} for linear activation functions and in \cite{pacelli2023statistical} for general activations,
the partition function can be evaluated analytically in the proportional-width limit, where $N_1,P\rightarrow\infty$ with $P/N_1=\alpha$ fixed.
The solution is obtained via a saddle-point evaluation of an integral, which corresponds to optimizing a scalar-dependent effective action $S(Q)$
through the condition
$S'(\bar{Q}) = 0$.

Importantly, the optimization depends on the training dataset (besides the temperature $T = 1/\beta$), so that $\bar{Q}$ contains information on both the input data
and their corresponding labels. In the following, we will compute average observables over the posterior Gibbs distribution $P(\theta|X,y)$ associated to $\mathcal{Z}$, denoted with
$\langle \star \rangle = \langle \star \rangle_{P(\theta|X,y)}$.

\emph{Collective displacement of the features---}
As shown in \cite{aiudi2023},
within this setting it is possible
to compute the posterior statistics of the features, which are collective variables of the first layer weights: $\sigma(h_j^\mu) = \sigma(\sum_{i_0}w_{ji_0}x_{i_0}^\mu/\sqrt{N_0})$. This is captured by the average similarity matrix, which reads \cite{aiudi2023}:
\begin{align} \label{SimMatrix}
    \langle \sigma(h_i^\mu)\sigma(h_i^\nu)\rangle &= K_{\mu\nu}
    - \underbrace{\frac{\bar{Q}}{\lambda_1N_1} \sum_{\lambda,\delta=1}^P K_{\mu\lambda} K_{\nu\delta}
    (\tilde{K}^{-1}_{(R)})_{\lambda\delta}}_{\Delta_1^{\mu\nu}}
    \quad + \nonumber \\
    & + \underbrace{\frac{\bar{Q}}{\lambda_1N_1} \sum_{\lambda,\delta=1}^P K_{\mu\lambda} K_{\nu\delta}
    (\tilde{K}_{(R)}^{-1}y)_\lambda (\tilde{K}_{(R)}^{-1}y)_\delta}_{\Delta_2^{\mu\nu}},
\end{align}
The NNGP kernel $K_{\mu\nu}$ is defined as the averaged similarity matrix of the features over the prior distribution, $K_{\mu\nu} = \langle\sigma(h^\mu)\sigma(h^\nu)\rangle_{P(h|X)}$,
and we have defined $\tilde{K}_{(R)} = \mathds{1}/\beta+K_{(R)}$,
where $K_{(R)}=\bar{Q}K/\lambda_1$
is the renormalized kernel.
Both in the proportional and in the infinite-width limit, the posterior corrections $\Delta_1^{\mu\nu}$, $\Delta_2^{\mu\nu}$ are of order $\mathcal{O}(1/N_1)$, thus irrelevant
whenever $N_1 \to \infty$. While this fact is
evident in the infinite-width limit, where the sums involve a finite number of terms $P$,
it requires further explanation when $P\rightarrow\infty$.
Indeed, if \(\beta \to \infty\), then \(\tilde{K}_{(R)}^{-1}\) is
equal to $\lambda_1 K^{-1}/\bar{Q}$. As a result, the first sum gives $\Delta_{1}^{\mu\nu} = K_{\mu\nu}/N_1$,
and the second gives $\Delta_2^{\mu\nu}=(\lambda_1/\bar{Q})y_\mu y_\nu/N_1$. On the other hand, if \(\beta \to 0\), we have \(\tilde{K}_{(R)}^{-1} \sim \beta\mathds{1} \to 0\), which makes both the terms in the sum vanish. For all intermediate values of \(\beta\), the system continuously interpolates between these two regimes
and the two terms will remain of order $\mathcal{O}(1/N_1)$.

These considerations point to the fact that
the second-order statistics of the features in the proportional-width limit is the same as in the
infinite-width limit, where no feature learning happens \cite{Neal,JacotNTK}.
We now show that it is still possible
to define observables that turn the subleading terms $\Delta_1$ and $\Delta_2$ into finite corrections in the proportional limit.
While the previous results hold for a regression task with general training labels, in what follows we consider binary labels $y^\mu = \pm 1$, and training datasets where the data points are equally split into the two different classes of cardinality $P/2$ (the generalization to unbalanced datasets \cite{pmlr-v202-francazi23a, pezzicoli2025anomaly} and other pairs of label values is shown in Sec.~S1 of the Supplemental Material (SM) \cite{SM}).

The squared distance $D^2$, defined as the separation between the mean post-activations of the two different classes, has been used in \cite{Ciceri2024} as a measure of class manifold separation between features:
\begin{equation}\label{eq:defD2}
    D^2 = \left\|\frac{2}{P}\sum_{\mu=1}^{P/2}\sigma(\bm{h}^{\mu}_+)-\frac{2}{P}\sum_{\mu=1}^{P/2}\sigma(\bm{h}^{\mu}_-)\right\|^2,
\end{equation}
where $\bm{h}_\pm^\mu$ is the pre-activation corresponding to the $\mu$-th sample of the first and second class, respectively.
Averaging over the posterior, we obtain:
\begin{align}
        \langle D^2\rangle &= \frac{4N_1}{P^{2}}\sum_{\mu,\nu=1}^{P}y_\mu y_\nu \langle\sigma(h_i^\mu)\sigma( h_i^\nu)\rangle = \nonumber\\
        & = \frac{4}{\alpha P}y^TKy
        - \frac{4}{\alpha P} y^T\Delta_1y +  \frac{4}{\alpha P}y^T\Delta_2y\, ,
        \label{eq:avgdist}
\end{align}
with $\Delta_1$ and $\Delta_2$ as in Eq.~\eqref{SimMatrix}.
It is important to emphasize that the $y$'s appearing explicitly in this formula do not originate from the posterior distribution, but rather from the relative signs in the definition (\ref{eq:defD2}).
The labels stemming from the posterior average are instead embedded within $\Delta_1$ and $\Delta_2$.
In fact, averaging over the prior yields the first term alone, $\langle D^2\rangle_{P(\theta)}=4y^TKy/(\alpha P)$.

In the proportional limit, the scaling of the three terms in Eq.~\eqref{eq:avgdist} is nontrivial,
but it can be discussed in the zero-temperature limit $\beta\rightarrow\infty$, where the expressions simplify. In this case, the second term reduces to $4 y^T K y/(\alpha PN_1)$, while the third simplifies to $4\lambda_1/\bar{Q}$.
The scaling of the first and second terms with $P$ is controlled by $y^TKy/P$. For iid Gaussian data, this ratio is provably $\mathcal{O}(1)$: since the entries of $K$ are independent random variables, the central limit theorem gives $\sum_{\nu}^PK_{\mu\nu}y_\nu\sim\sqrt{P}$, so that the double sum $y^TKy\sim P$. In this case, the second term vanishes in the proportional limit, and the last term alone provides a finite, feature-dependent correction to the distance averaged over the prior.
For real datasets, this scaling is not guaranteed, and $y^TKy/P$ need not remain $\mathcal{O}(1)$: if the kernel entries are correlated with the label structure of the dataset, this correlation survives the sum over $\mu,\nu$ instead of averaging out, and $y^TKy/P$ can grow as $\mathcal{O}(P)$ rather than $\mathcal{O}(1)$. Therefore, we do not drop the second term when working with real data. However, for the datasets considered in this work (e.g., CIFAR-10, see Fig.~S1 of the SM \cite{SM}), we find numerically that, while the second term remains finite and non-negligible in the large but finite-size regime where the theory is applicable \cite{baglioni2024predictive}, it stays small relative to the third term, which is the one responsible for the behavior we discuss below. In Sec.~S1 of the SM \cite{SM}, we provide a numerical characterization of this scaling for specific datasets and splits (Fig.~S1), together with a discussion of situations where the prior term dominates the distance (Fig.~S2). Keeping all three contributions, the squared distance reads:
\begin{align}
    & \langle D^2\rangle = \frac{4}{\alpha P}y^TKy
        - \frac{4}{\alpha P} y^T\Delta_1y +
    \frac{\lambda_1}{\bar{Q}}(\bar f_+-\bar f_-)^2,
    \label{eq:D2_approx}\\
    &  \bar f _{\pm} = \frac{2}{P}\sum_{\mu=1}^{P/2} \langle f_\pm^\mu\rangle ,
    \hspace{0.3cm}
    \langle f_\pm^\mu \rangle = \sum_{\lambda,\sigma=1}^{P}(K_{(R)})_{\mu\lambda}(\tilde{K}^{-1}_{(R)})_{\lambda\sigma}y_\sigma. \nonumber
\end{align}
where $\langle f_{\pm}^\mu\rangle$ are the posterior expected outputs of the NN in the proportional limit \cite{pacelli2023statistical, pacelli2024GP}, and $\Delta_1$ is as in Eq.~\eqref{SimMatrix}. The symbols $\pm$ indicate that the index $\mu$ runs over the kernel evaluated on samples from the first and second class, respectively.

Note that the squared distance
diverges in the infinite-width limit $\alpha\rightarrow0$,
since the post activations $\sigma(\bm{h})$ are $N_1$-dimensional vectors, and their squared norms are $\mathcal{O}(N_1)$.
We plot $\left<D^2\right>$ as a function of the temperature in Fig.~\ref{fig:distance}(b)
(and $\alpha\left<D^2\right>$ in Fig.~\ref{fig:distance}(c), which is rescaled so that it converges in the infinite-width limit).
The dependence of the squared distance on the temperature is nonmonotonic for $\alpha>0$, with a characteristic peak at positive temperature.
(An alternative way to normalize the distance is presented in Sec.~S1 of the SM \cite{SM}, with similar results as here.)
The emergence of the term $(\bar f_+-\bar f_-)^2 \lambda_1/\bar{Q}- 4 y^T\Delta_1y/(\alpha P)$ in the proportional limit signals the network’s ability to learn the features by exploiting contributions from the posterior, which influences their separation.
Thus, the strongly nonmonotonic behavior of $\left<D^2\right>$ is understood as a consequence of feature learning. This is confirmed by the fact that assigning random labels to the data nearly removes the peak (dashed line in Fig.~\ref{fig:distance}(b)).
The same phenomenology is observed for other class splits of the CIFAR-10 and MNIST datasets (Fig.~S3 of the SM \cite{SM}).
Furthermore, reducing $\alpha$ towards the infinite-width limit reduces the relative contribution of the posterior with respect to the prior, while simultaneously eliminating the peak (Fig.~\ref{fig:distance}(c)).
This is in accordance with the fact that the posterior statistics becomes indistinguishable from that of the prior in the infinite-width limit.
Note that while temperature can act like a regularizer, the observed non-monotonicity in manifold distance is not purely a regularization effect. Unlike ridge-like regularization, which modifies outputs even for fixed features (e.g., in the infinite-width limit), temperature shapes the geometry of the representations themselves in a way that reflects task-relevant semantic structure, as the peak is absent when labels are randomized.

\begin{figure}[t]
  \centering
  \includegraphics[width=0.4\textwidth]{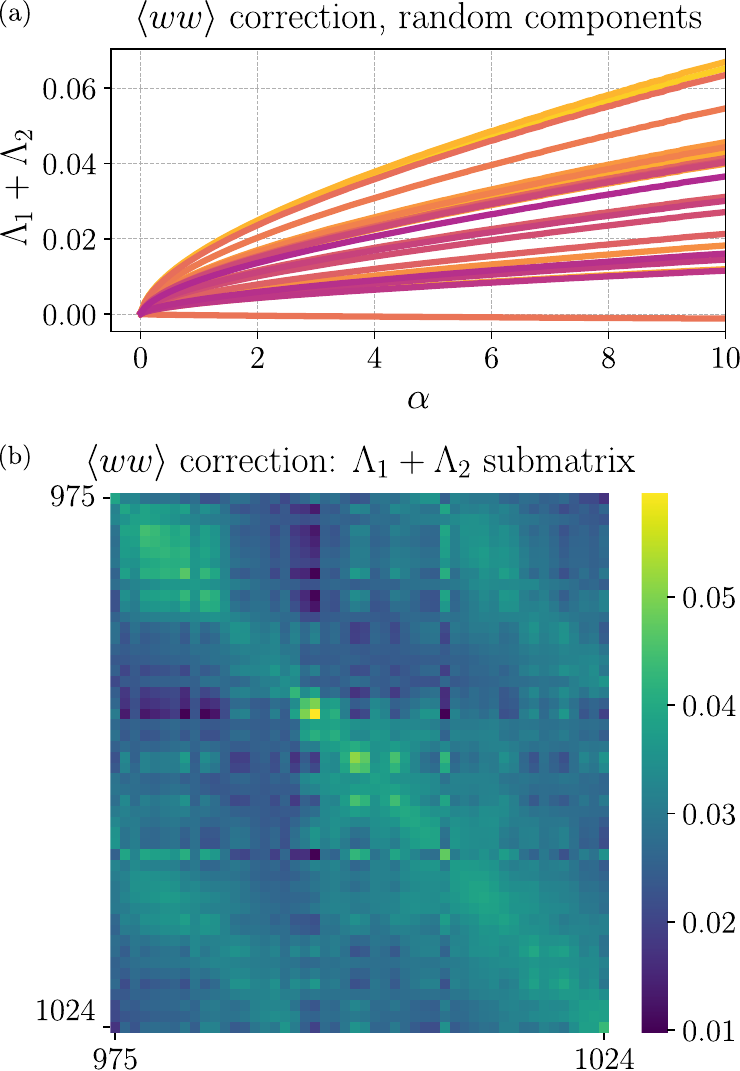}
  \caption{\textbf{Finite corrections to $\langle ww\rangle$ in the proportional regime.} (a) Values of 25 randomly selected components of the proportional correction $\Lambda_1+\Lambda_2$, Eq.~\eqref{eq:lambda_2}, as a function of $\alpha$. As $\alpha$ increases, transitioning the system to the proportional regime, each component exhibits a linear trend, consistent with the scaling predicted by GD dynamics. (b) The heatmap shows a submatrix of $\Lambda_1+\Lambda_2$ at $\alpha = 2$. The magnitude of the matrix elements is not negligible compared to the value $\langle ww\rangle_{ij} = \delta_{ij}/\lambda_0$ expected at infinite width (here, $\lambda_0=1$).
  Simulations were performed with 1500 samples of CIFAR10 dataset, split into even and odd classes. We used the activation function $\sigma = \text{erf}$ and $T=10^{-3}$.}
  \label{fig:correlations}
\end{figure}

\emph{Microscopic pairwise correlations of the weights---}
In the previous section, we isolated a \emph{macroscopic} observable that carries information about feature learning at proportional width. Here, we investigate \emph{microscopic} features, showing that
the statistics of the first-layer weights $w$ depends on the data (which does not happen in the infinite-width limit).
We compute the two-point function $ \langle w_{1k}w_{1h}\rangle$ with respect to the posterior distribution.
In Sec.~S2 of the SM \cite{SM}, we show that it is possible to use the effective action formalism to express the similarity matrix $\langle w_{1h}w_{1k}\rangle$ in terms of the order parameter $\bar{Q}$:
\begin{align}
    &\langle w_{1h}w_{1k}\rangle = \frac{\delta_{hk}}{\lambda_0} + \frac{(\Lambda_1)_{hk}}{\lambda_0} + \frac{(\Lambda_2)_{hk}}{\lambda_0},\label{eq:ww}\\
    &\, \, (\Lambda_1)_{hk} = \frac{\alpha\bar{Q}}{2\lambda_1}\frac{1}{P}
    \sum_{\mu,\nu=1}^{P}
(\tilde{K}^{-1}_{(R)})_{\mu\nu}\Delta_{hk}^{\mu\nu}, \label{eq:lambda_1} \\
    & \, \, (\Lambda_2)_{hk} = \frac{\alpha\bar{Q}}{2\lambda_1}\frac{1}{P}\sum_{\mu,\nu=1}^P(\tilde{K}^{-1}_{(R)}y)_\mu(\Delta_{hk}^{\mu\nu})
    (\tilde{K}^{-1}_{(R)}y)_\nu. \label{eq:lambda_2}
\end{align}
$\Delta_{hk}^{\mu\nu}$ is a data-dependent kernel defined in Sec.~S2 of the SM \cite{SM}.
In the case of a linear activation function ($\sigma=\mathrm{Id}$), these results hold exactly for finite $N_1$ and $P$, and Eq.~\eqref{eq:ww} reproduces Eq.~\eqref{SimMatrix}, as we verify in Sec.~S3 of the SM \cite{SM}.
This fact is in line with the use of the effective action formalism carried out in \cite{JMLR:v26:24-1158} for finite-width deep linear networks.

In the infinite-width limit, the terms $\Lambda_1$ and $\Lambda_2$ vanish,
the components of $w$ become uncorrelated, and averages over the prior and over the posterior coincide.
In the proportional case, since the kernel $\Delta_{hk}^{\mu\nu}$ cannot be expressed in simple terms by using $K_{\mu\nu}$, the zero-temperature limit is not useful to investigate the magnitude of the proportional correction terms.
In Fig.~\ref{fig:correlations}, we show results of numerical simulations performed with $\sigma = \text{erf}$. In this case, the kernel is expressed through the data as $\Delta_{hk}^{\mu\nu} = -({x_h^\mu x_k^\nu + x_h^\nu x_k^\mu})/(\lambda_0N_0)\cdot 4/\pi(\text{det}(1+2C))^{-1/2}$, where $C$ is the empirical covariance between the couple $(\mu,\nu)$ of datapoints. We note that the proportional correction $\Lambda_1+\Lambda_2$ contributes to the statistics of the weights with a finite term that depends almost linearly on $\alpha$.
The linearity is consistent with a scaling argument that can be invoked also in the context of GD dynamics: in this case, the time derivative of the weights is proportional to $\sqrt{\alpha}$.

\emph{Discussion and conclusions---}
These results contribute to the debate around the definition, interpretation, and theoretical analysis of feature learning. The relevance is twofold. First, we propose the temperature-dependent interclass separation as a macroscopic observable that signals the departure from the infinite-width limit, and depends on the task labels.
Second, we show that Bayesian networks in the proportional limit develop internal features with nontrivial statistics, already in the two-layer architecture that we investigate, despite the fact that the posterior process remains a Gaussian process with a trivially rescaled kernel, even in multilayer models \cite{baglioni2026kernelrenormalizationbayesiandeep}: the posterior mean predictor retains the functional form of the prior (infinite-width) predictor, up to an effective renormalization of the temperature. The takeaway is that feature learning in these architectures is entirely encoded in the internal (microscopic and collective) observables, rather than in the learned mean predictor, whose dependence on temperature is already fixed at infinite width.

The squared distance $D^2$ is a collective observable of the features
that signals non-trivial feature learning in Bayesian FC shallow networks.
Other measures of feature segregation have been found to display interesting behavior empirically \cite{FrosstPapernot:2019, HeSu:2023, springblock};
the Bayesian proportional-width framework employed here may be able to capture those as well.

Our computations show that the posterior average $\langle D^2\rangle$ is a nonmonotonic function of the temperature $T$.
The peak at the inversion point is more pronounced for larger values of $\alpha$, and eventually disappears in the infinite-width limit $\alpha=0$.
A possible interpretation of this behavior arises from the observation that increasing the temperature allows the posterior to sample
regions of the loss landscape further away from the optima.
Intuitively, gradient descent dynamics visits these regions earlier
than those with smaller loss
(in simpler settings, it can be proved that temperature acts as
early stopping in the optimization dynamics \cite{pmlr-v89-ali19a}).
The behavior of the distance then aligns with the inversion dynamics reported in \cite{Ciceri2024}, where the
nonmonotonicity of $D^2$ appears as a function of the training error.
This behavior has been interpreted as a trade-off between
the segregation of the two class manifolds and the fine tuning required for classifying the last hard samples.
In turn, the transition between the easy and hard phases of training
has been interpreted as a manifestation of feature learning.
The results summarized in Fig.~\ref{fig:distance} may be seen as the equilibrium counterpart to this non-equilibrium phenomenon.

The (posterior) second-order statistics of the hidden-layer weights
develop finite correlations in the proportional limit.
This is remarkable.
The Gaussian process of the output receives a trivial modification with respect to the infinite-width limit, because the scalar renormalization by $\bar{Q}$ can be reabsorbed in the prior parameter $\lambda_1$ \cite{aiudi2023}.
In contrast, the second-order statistics of $w$ depends on the input patterns in a way that cannot be traced back to the infinite-width limit.

Finally, we would like to point out two distinctions between our work and other lines of research in this field.
First,
our results hold for standard-scaled neural networks. Another possibility would be to consider the mean-field scaling, where microscopic quantifiers of feature learning have been found already in the infinite-width limit \cite{tensoriv, seroussi2023natcomm, fischer24critical, rubin2025from,bordelon2022selfconsistent, lauditi2025adaptive}.
Second,
the proportional regime represents a genuinely overparameterized scenario. Therefore, our theory does not help to characterize feature learning closer to the interpolation threshold. This setting, which may also be relevant for modern deep learning applications, has been very recently investigated for Gaussian training inputs \cite{lopezpolinomial2025, barbier2025optimal,barbier2025extensive,erba2025quadratic}, leading to more complex feature learning mechanisms than the ones under scrutiny here \cite{ingrossoConv, doi:10.1073/pnas.2311805121, ringel2025applications}.

\begin{acknowledgments}
P.R.~is supported by $\#$NEXTGENERATIONEU (NGEU) and funded by the Ministry of University and Research (MUR), National Recovery and Resilience Plan (NRRP), project MNESYS (PE0000006) ``A Multiscale integrated approach to the study of the nervous system in health and disease'' (DN. 1553 11.10.2022). R.P.~is funded by the MUR, project PRIN 2022HSKLK9.
\end{acknowledgments}

\bibliography{biblio}

\clearpage
\onecolumngrid
\begin{center}
\textbf{\large Supplemental Material}
\end{center}

\setcounter{section}{0}
\setcounter{equation}{0}
\setcounter{figure}{0}
\renewcommand{\thesection}{S\arabic{section}}
\renewcommand{\theequation}{S\arabic{equation}}
\renewcommand{\thefigure}{S\arabic{figure}}
\renewcommand{\theHsection}{S\arabic{section}}
\renewcommand{\theHequation}{S\arabic{equation}}
\renewcommand{\theHfigure}{S\arabic{figure}}

\section{Derivation of the average squared distance}\label{sec:distance}

\begin{figure}[h!]
    \centering
    \includegraphics[width=0.8\textwidth]{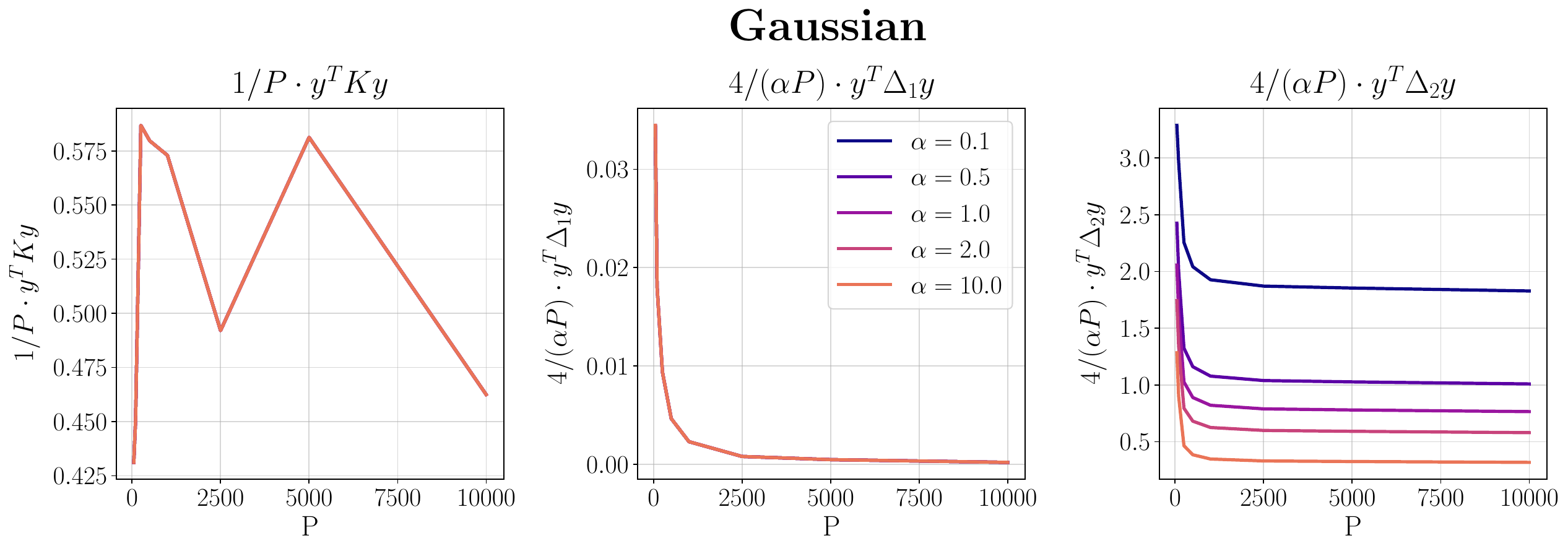}\\[-0.15cm]
    \includegraphics[width=0.8\textwidth]{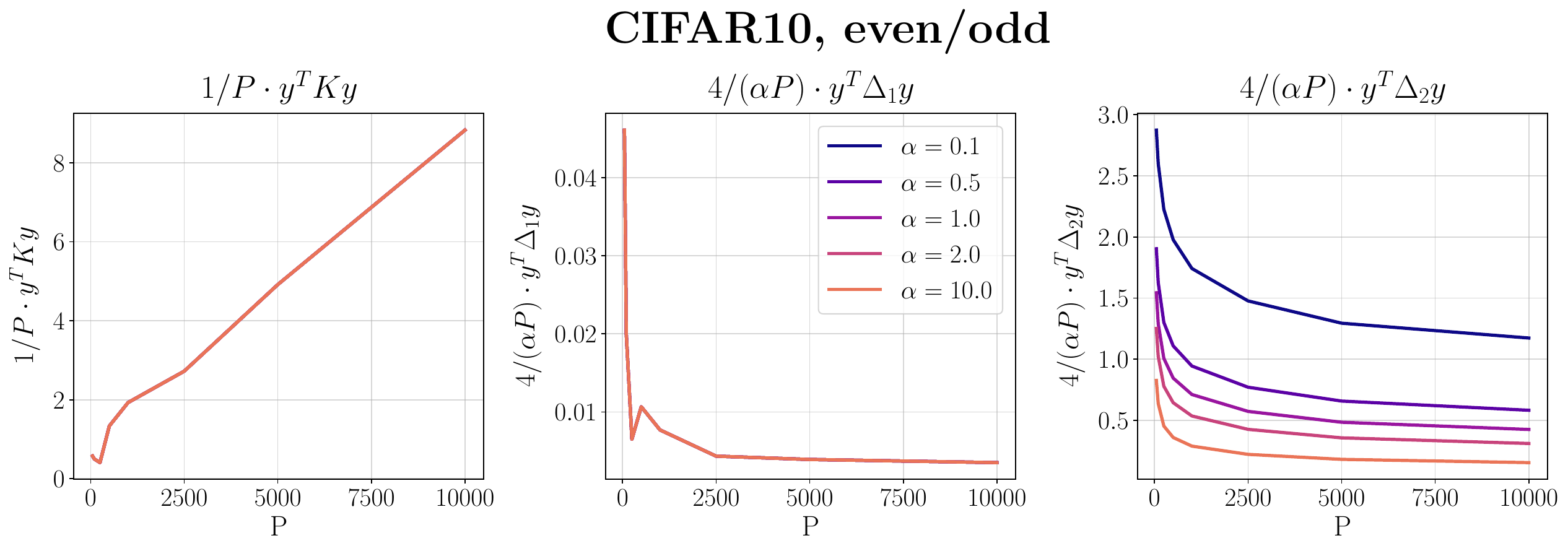}\\[-0.15cm]
    \includegraphics[width=0.8\textwidth]{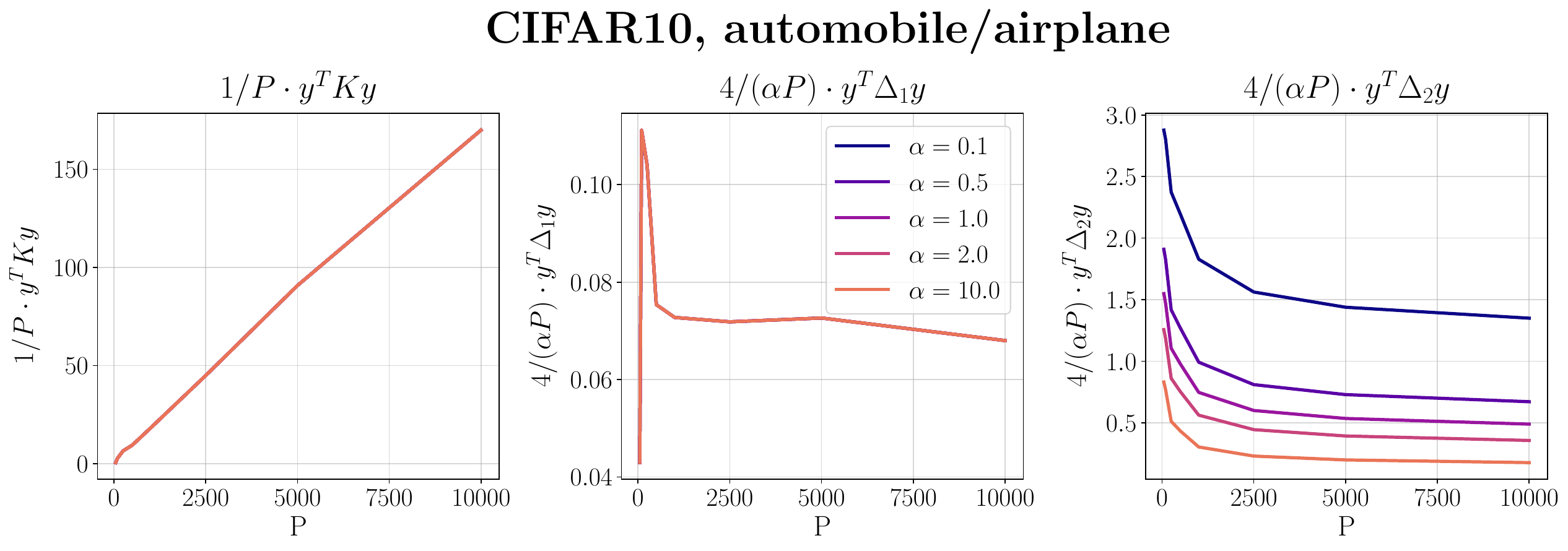}    
    \includegraphics[width=0.8\textwidth]{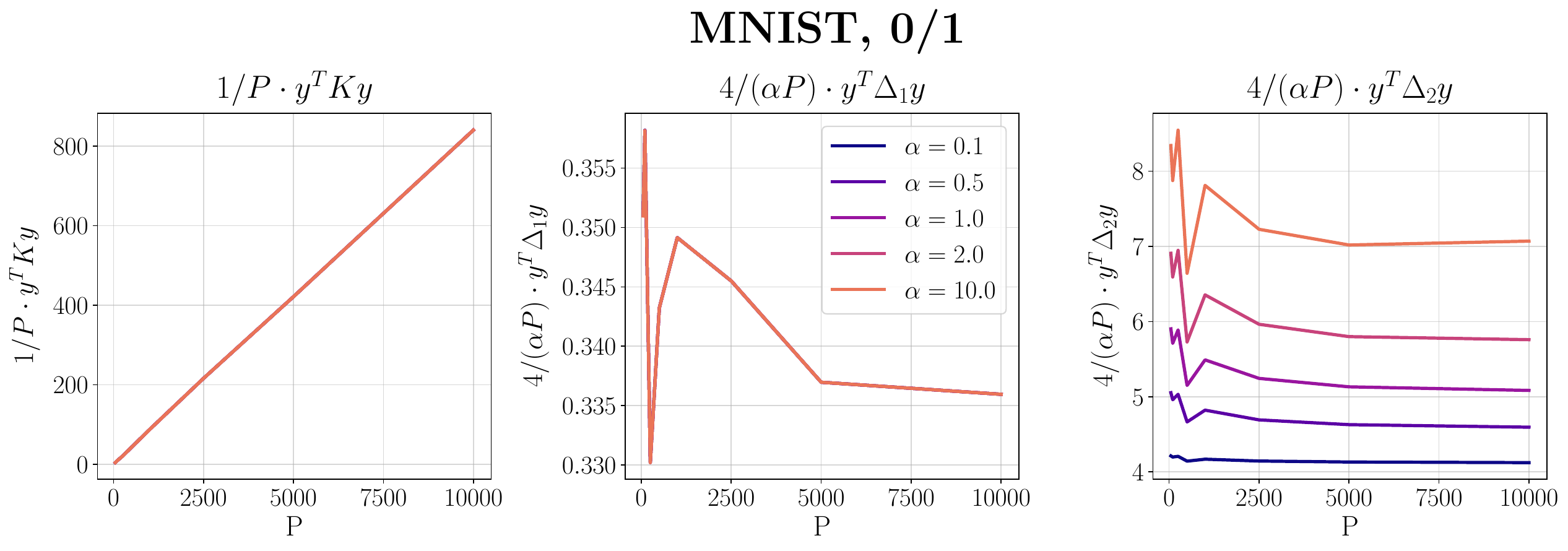}
    \caption{Gaussian data are associated with the theoretical scaling provided in the text. The scaling is not satisfied for CIFAR10 dataset, where $z^TKz/P$ grows linearly (albeit very slowly). However, since the effective action formalism has been proven to work for large but finite $P$ and $N_1$, the central point is to identify a regime where the magnitude of $\tilde{D}_1^2$ is similar to the one of $\tilde{D}_3^2$. In this case the first term grows slowly, allowing us to compare the two contributions. This approach cannot be pursued in the case of MNIST dataset restricted to $0$ and $1$ classes. In this case, the kernel strongly correlates with the labels (see Fig.~\ref{fig:kernel}), highlighting a linear scaling with a more pronounced growth rate. The graphs are plotted for the temperature value $T=0.001$.}
    \label{fig:scaling}
\end{figure}

In the main text, we mentioned that the squared distance can be defined to be finite also in the infinite-width regime and in the case of unbalanced datasets, where each class has a different number of data points. A more general definition can be considered:
\begin{equation}
    D^2 = \frac{1}{N_1^\eta}\left\|\frac{1}{N_+^\delta}\sum_{\mu=1}^{N_+}\sigma(\bm{h}_+^{\mu})-\frac{1}{N_-^\delta}\sum_{\mu=1}^{N_-}\sigma(\bm{h}_-^{\mu})\right\|^2,
\end{equation}
where $\eta \geq 0$, $\delta > 0$, and $N_+$ and $N_-$ (with $N_+ + N_- = P$) denote the number of training data points belonging to the first and second class, respectively. Assuming that $N_+$ and $N_-$ represent non-negligible fractions of the total number of points $P$ --- which is large in the proportional limit --- we can define $\gamma = \left( \frac{N_+}{N_-} \right)^\delta$ and express the distance as
\begin{equation}
    D^2 = \frac{1}{N_1^\eta N_+^{2\delta}}
    \begin{bmatrix}
        \sigma(\bm{h}_+) \hspace{0.15cm} \sigma(\bm{h}_-)
        \end{bmatrix}
        \cdot
        \begin{bmatrix}
        1 & -\gamma \\ 
        -\gamma & \gamma^2
        \end{bmatrix}
        \cdot
        \begin{bmatrix}
        \sigma(\bm{h}_+) \\ 
        \sigma(\bm{h}_-)
    \end{bmatrix},
\end{equation}
where the matrix and the vectors are to be understood in block form. By defining the vector $z$, with components
\[
z_\mu = 
\begin{cases}
1 \hspace{1.3cm}\text{for}\hspace{0.2cm}  \mu=1,...,N_+\\
-\gamma \hspace{1cm}\text{for}\hspace{0.2cm}  \mu=N_++1,...,P
\end{cases}\hspace{0.1cm},
\]
it is possible to write the block matrix as a tensor product whose components are $z_\mu z_\nu$, which allows writing the squared distance as
\begin{equation}
    D^2 = \frac{1}{N_1^\eta N_+^{2\delta}}\sum_{\mu\nu}^{P}z_\mu z_\nu\sum_{i}^{N_1}\sigma(h_i^\mu)\sigma( h_i^\nu) . 
\end{equation}
Note that in the case of a balanced dataset with $N_+ = N_-$, we have $\gamma=1$ and $z_\mu=\pm 1$, which are the values assumed by the labels $y_\mu$, as in the case of the main text. Averaging over the posterior distribution, we have:
\begin{equation}
    \langle D^2\rangle = \frac{1}{N_1^\eta N_+^{2\delta}}\sum_{\mu\nu}^{P}z_\mu z_\nu\sum_{i}^{N_1}\langle \sigma(h_i^\mu)\sigma(h_i^\nu)\rangle.
\end{equation}
The averaged similarity matrix has been computed in \cite{aiudi2023}, which provides an explicit expression for the averaged similarity matrix: 
\begin{equation}\label{SimilarityMatrix}
    \langle \sigma(h_i^\mu) \sigma(h_i^\nu)\rangle = K_{\mu\nu}-\frac{\bar{Q}}{\lambda_1N_1}\sum_{\lambda\delta}K_{\mu\lambda}K_{\nu\delta}\left[(\tilde{K}^{-1}_{(R)})_{\lambda\delta}-(\tilde{K}_{(R)}^{-1}y)_\lambda(\tilde{K}_{(R)}^{-1}y)_\delta\right].
\end{equation}
Since the averaged similarity matrix does not depend on the index $i$ we can trivially perform the internal sum to get:
\begin{equation}
        \langle D^2\rangle = \frac{1}{N_1^{\eta-1} N_+^{2\delta}}\sum_{\mu\nu}^{P}z_\mu z_\nu\langle \sigma(h_i^\mu) \sigma(h_i^\nu)\rangle,
\end{equation}
which can be expressed as
\begin{align}
    \langle D^2\rangle &= \frac{1}{N_1^{\eta-1} N_+^{2\delta}}\sum_{\mu\nu}^{P}z_\mu z_\nu\left[K_{\mu\nu}-\frac{\bar{Q}}{\lambda_1N_1}\sum_{\lambda\delta}K_{\mu\lambda}K_{\nu\delta}\left[(\tilde{K}^{-1}_{(R)})_{\lambda\delta}-(\tilde{K}_{(R)}^{-1}y)_\lambda(\tilde{K}_{(R)}^{-1}y)_\delta\right]\right] = \nonumber \\ &
    =    \tilde{D}^2_1 - \tilde{D}^2_2 + \tilde{D}^2_3.
\end{align}
As mentioned in the main text, the terms that come from the posterior distribution (which involve $y$) are clearly distinguished from the ones which come from the relative minus sign of the norm (represented by $z$). We defined
\begin{equation}
    \tilde{D}_1^2 = \frac{1}{N_1^{\eta-1} N_+^{2\delta}}\sum_{\mu\nu}^{P}z_\mu  K_{\mu\nu} z_\nu,
\end{equation}
\begin{equation}
    \tilde{D}_2^2 = \frac{\bar{Q}}{\lambda_1}\frac{1}{N_1^{\eta} N_+^{2\delta}}\sum_{\mu\nu}^{P}z_\mu z_\nu \sum_{\lambda\delta}K_{\mu\lambda}K_{\nu\delta}(\tilde{K}^{-1}_{(R)})_{\lambda\delta},
\end{equation}
\begin{align}
    \tilde{D}_3^2 &= \frac{\bar{Q}}{\lambda_1}\frac{1}{N_1^{\eta} N_+^{2\delta}}\sum_{\mu\nu}^{P}z_\mu z_\nu 
    \sum_{\lambda\delta}K_{\mu\lambda}K_{\nu\delta}\sum_{\sigma\rho}(\tilde{K}^{-1}_{(R)})_{\lambda\sigma}
    (\tilde{K}^{-1}_{(R)})_{\delta\rho}y_\sigma y_\rho = \nonumber \\
    &=\frac{\bar{Q}}{\lambda_1}\frac{1}{N_1^{\eta} N_+^{2\delta}}\sum_{\mu\nu}^{P}z_\mu z_\nu \left(\sum_{\lambda\sigma}K_{\mu\lambda}(\tilde{K}^{-1}_{(R)})_{\lambda\sigma}y_\sigma\right)\left(\sum_{\delta\rho}K_{\nu\delta}(\tilde{K}^{-1}_{(R)})_{\delta\rho}y_\rho\right) = \nonumber \\
    &= \frac{\lambda_1}{\bar{Q}}\frac{1}{N_1^{\eta} N_+^{2\delta}}\left(\sum_{\mu}^{P}z_\mu \langle f\rangle_\mu\right)^2,
\end{align}
where $\langle f_\mu \rangle$ is defined as in the main text $\langle f_\mu \rangle= \sum_{\lambda,\sigma=1}^{P}(K_{(R)})_{\mu\lambda}(\tilde{K}^{-1}_{(R)})_{\lambda\sigma}y_\sigma$ and represents the expected output of the Gaussian Process associated with the Neural Network in the proportional limit.
It can be observed that the expected output, although defined through a sum involving an increasing number of terms in the proportional limit, has finite size. In fact, if $\beta \rightarrow \infty$, then $\tilde{K}_{(R)}^{-1}$ becomes proportional to $K^{-1}$. The result is thus a delta, and the output satisfies $\langle f_\mu \rangle = y_\mu$. On the other hand, if $\beta \rightarrow 0$, then $\tilde{K}_{(R)}^{-1} \sim \beta \mathds{1} \rightarrow 0$, which drives the outputs to zero. One can numerically check that all intermediate values of $\beta$ smoothly interpolate between these two regimes, while keeping $\langle f_\mu\rangle$ of finite size. With this in place, we can proceed further by writing
\begin{align}
    \sum_{\mu}^Pz_\mu \langle f_\mu \rangle & = \sum_{\mu}^{N_+}\langle f_+^{\mu}\rangle  -\gamma\sum_{\mu}^{N_-} \langle f_-^{\mu}\rangle= \nonumber \\
    & = N_+\left(\bar{f}_+-\gamma^{\frac{\delta-1}{\delta}}\bar{f}_-\right).
\end{align}
We defined $\bar{f}_\pm^{\mu} = 1/N_{\pm}\sum_{\mu}\langle f_\pm^{\mu}\rangle$. Since in the proportional limit all the quantities $N_1, P, N_+, N_- \rightarrow \infty$ at the same rate, it is useful to express everything as a function of $P$. From the relations $N_1 = P/\alpha$ and
\begin{equation}
    P = \left(1 + \frac{1}{\sqrt[\delta]{\gamma}}\right) N_+ = \tilde{\gamma} N_+ ,
\end{equation}
we can explicitly write the three contributions to the distance:
\begin{equation}
    \tilde{D}_1^2 = \frac{\alpha^{\eta-1}\tilde{\gamma}^{2\delta}}{P^{\eta+2\delta-1}}\sum_{\mu\nu}^{P}z_\mu K_{\mu\nu} z_\nu ,
\end{equation}
\begin{equation}
    \tilde{D}_2^2 = \frac{\bar{Q}}{\lambda_1}\frac{\alpha^{\eta}\tilde{\gamma}^{2\delta}}{P^{\eta+2\delta}}\sum_{\mu\nu}^{P}z_\mu z_\nu \sum_{\lambda\delta}K_{\mu\lambda}K_{\nu\delta}(\tilde{K}^{-1}_{(R)})_{\lambda\delta},
\end{equation}
\begin{equation}
    \tilde{D}_3^2 = \frac{\lambda_1\alpha^{\eta}\tilde{\gamma}^{2\delta-2}}{\bar{Q}}\frac{1}{P^{\eta+2\delta-2}}\left(\bar{f}_+-\gamma^{\frac{\delta-1}{\delta}}\bar{f}_-\right)^2.
\end{equation}
The zero temperature limit can be used to justify the different scalings in $P$ of the three quantities. When $T\rightarrow 0$, the internal sum of the second contribution returns $K_{\mu\nu}$, telling us that $\tilde{D}_2^2$ can be neglected. Under the assumption that $1/P\sum_{\mu\nu}z_\mu K_{\mu\nu} z_\nu \sim\mathcal{O}(1)$, the term $\tilde{D}_1^2$ is of the same order of $\tilde{D}_3^2$, making it a feature dependent correction that reflects the fact that the posterior is learning through the labels. As mentioned in the main text, the assumption is true in the case of Gaussian data, where the kernel has independent entries, which make the sum $\sum_\nu^P K_{\mu\nu}z_\nu \sim \sqrt{P}$ and $1/P\sum_{\mu\nu}z_\mu K_{\mu\nu}z_\nu = \mathcal{O}(1)$ in virtue of the central limit theorem. In the case of fairly noisy data, it is reasonable to expect the term to remain finite also for large values of $P$. As shown in Fig.~\ref{fig:scaling} ($\eta=0$, $\delta=1$), this is the case for the CIFAR10 dataset.

\begin{figure}[h]
    \centering

    \begin{minipage}{0.4\textwidth}
        \centering
        \includegraphics[width=\textwidth]{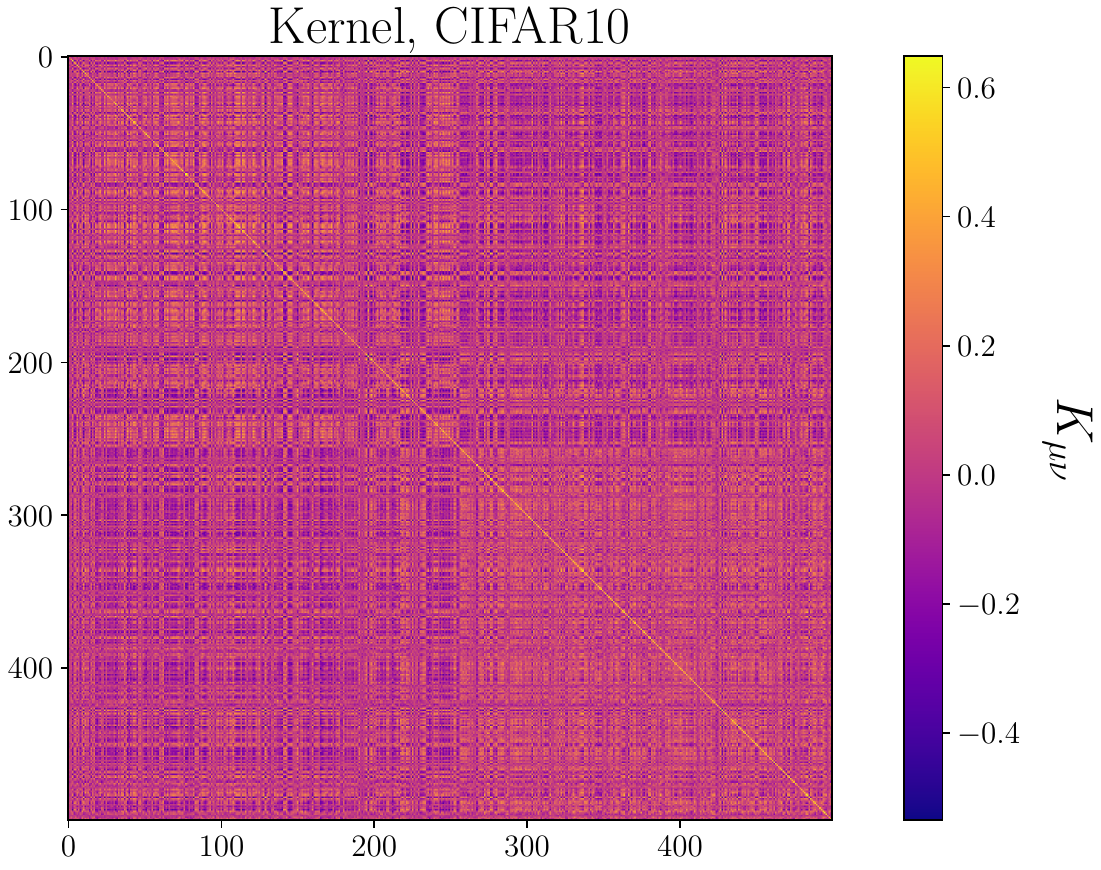}
    \end{minipage}
    \hspace{0.2cm}
    \begin{minipage}{0.4\textwidth}
        \centering
        \includegraphics[width=\textwidth]{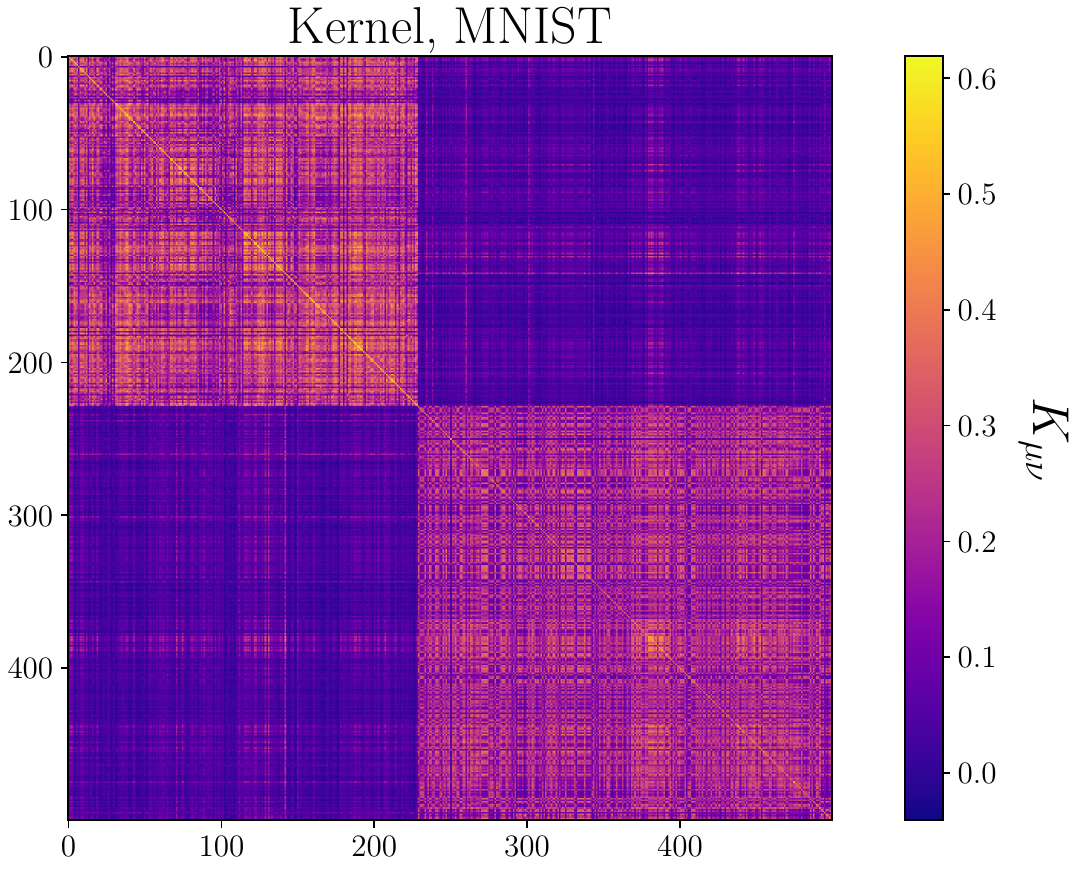}
    \end{minipage}

    \caption{
    The left panel shows the kernel components computed for $P=500$ CIFAR10 examples, split into even and odd classes, while the right panel shows the kernel components for $P=500$ MNIST examples, split into classes 0 and 1. For MNIST, the positive values of the kernel correlate with $z$, resulting in a linear scaling of $z^TKz/P$. For noisier datasets, this effect is still present, but the growth rate is strongly mitigated.
    }
    \label{fig:kernel}
\end{figure}

This type of argument holds for every choice of the scaling that meets the condition $\eta+2\delta-2=0$. One possible choice, which is the one presented in the main text, consists in selecting $\eta=0$, $\delta=1$, which returns (disregarding the second contribution):
\begin{equation}
    \langle D^2\rangle = \frac{1}{\alpha}\left(\frac{\gamma+1}{\gamma}\right)^2\frac{1}{P}\sum_{\mu\nu}^P z_\mu K_{\mu\nu} z_\nu  +
    \frac{\lambda_1}{\bar{Q}}(\bar{f}_+-\bar{f}_-)^2.
\end{equation}
As argued in the main text, the divergence in the infinite-width limit $\alpha\rightarrow 0$ is completely understood in terms of the original definition of the distance, which diverges as the norm of a vector with a growing number of components. As an alternative, the definition with $\eta=1$ and $\delta=1/2$ returns a well-defined infinite-width limit where the distance reads:
\begin{equation}
     \langle D^2\rangle = \frac{\gamma^2+1}{\gamma^2}\frac{1}{P}\sum_{\mu\nu}z_\mu K_{\mu\nu} z_\nu  + \alpha\frac{\lambda_1}{\bar{Q}}\frac{\gamma^2}{\gamma^2+1}\left(\bar{f}_+-\frac{1}{\gamma}\bar{f}_-\right)^2
 \end{equation}
 Note that in both cases the relative magnitude between the prior term and the posterior correction is proportional to $\alpha$. Because of this, in the infinite-width limit the prior term is dominant and the posterior correction can be disregarded, returning a distance that does not depend on the labels.

\begin{figure}[h]
    \centering

    \includegraphics[width=0.3\textwidth]{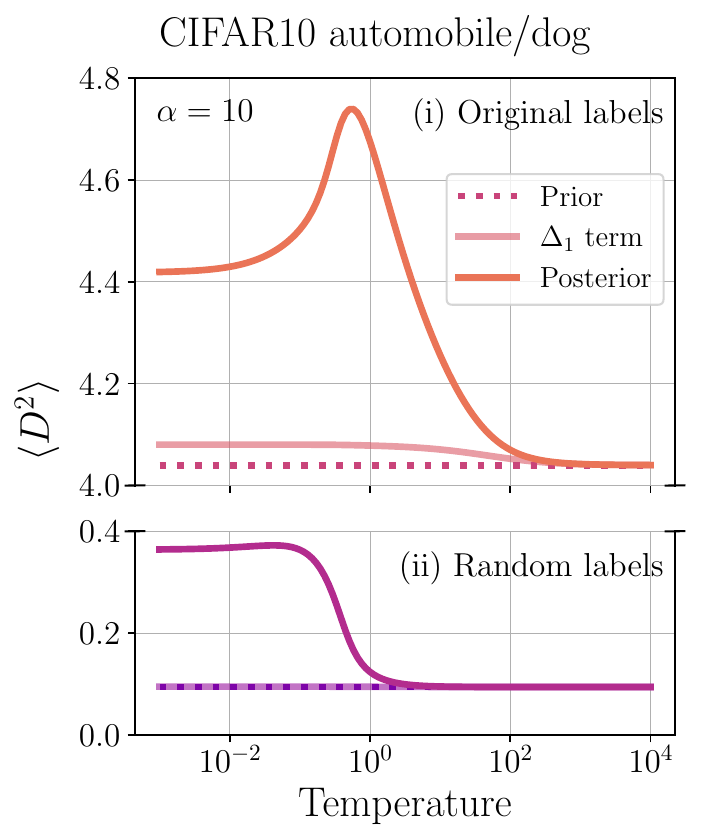}
    \includegraphics[width=0.3\textwidth]{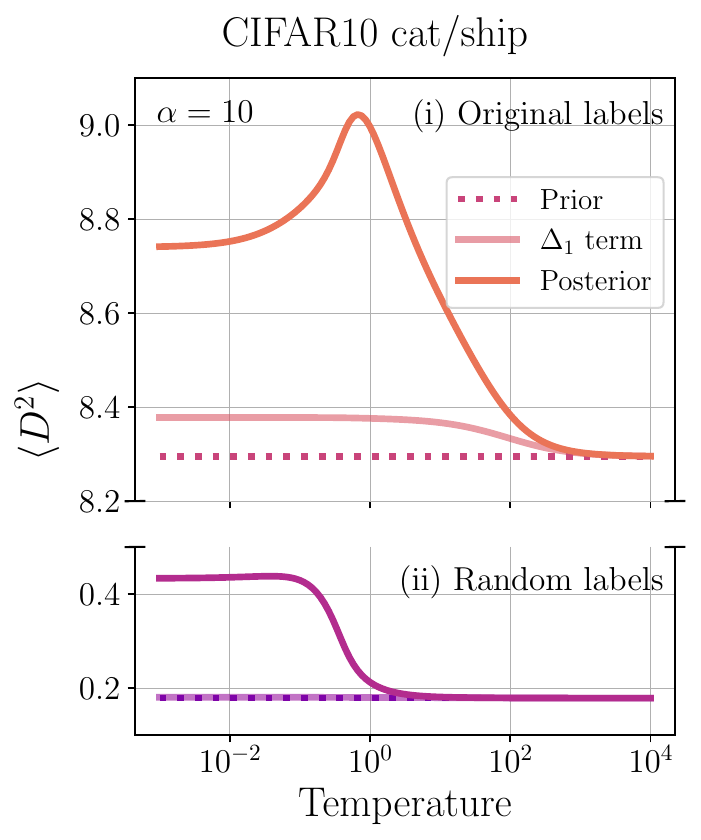}    
    \includegraphics[width=0.3\textwidth]{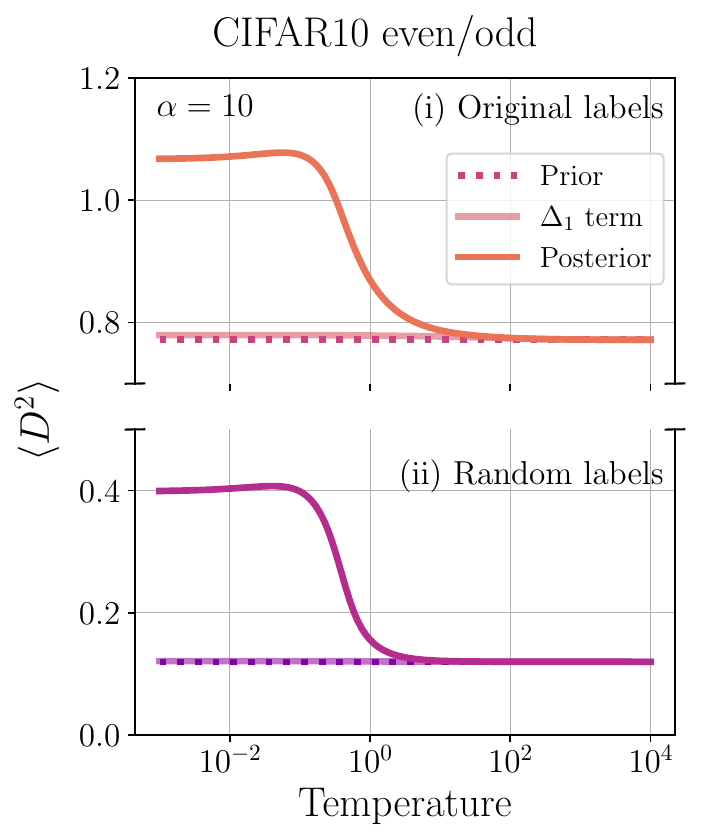}

    \vspace{0.5em}

    \includegraphics[width=0.3\textwidth]{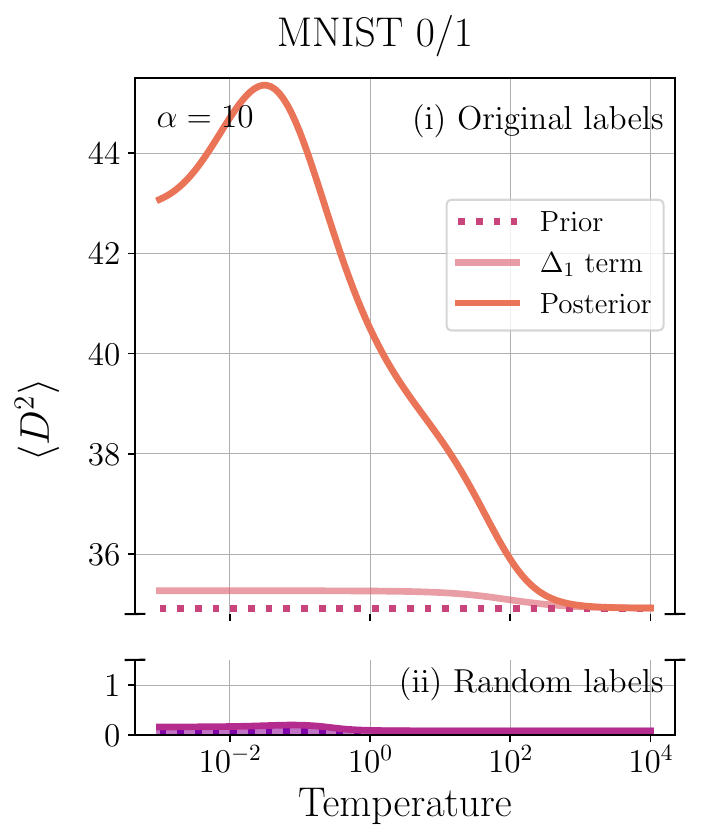}
    \includegraphics[width=0.3\textwidth]{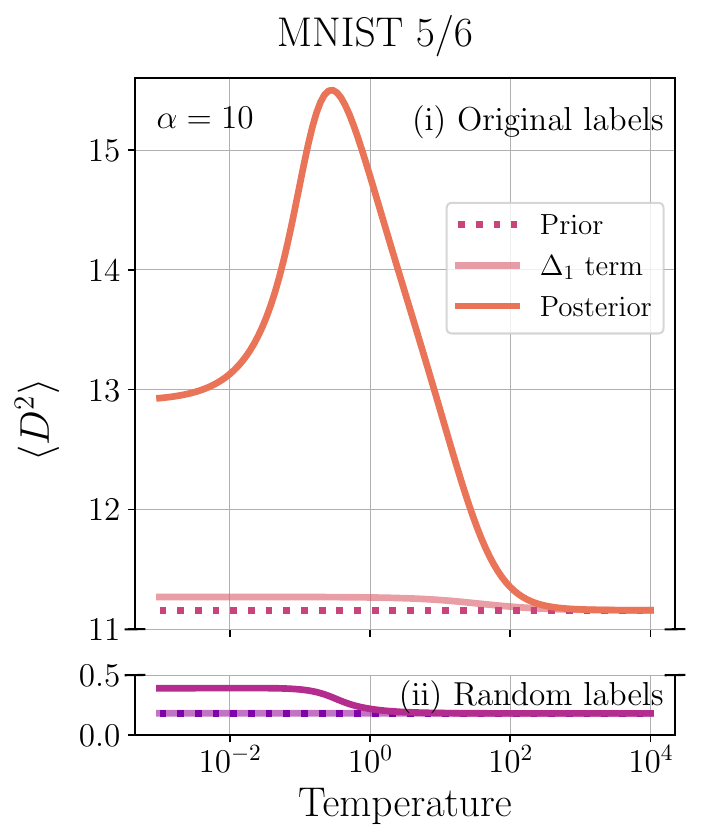}
    \includegraphics[width=0.3\textwidth]{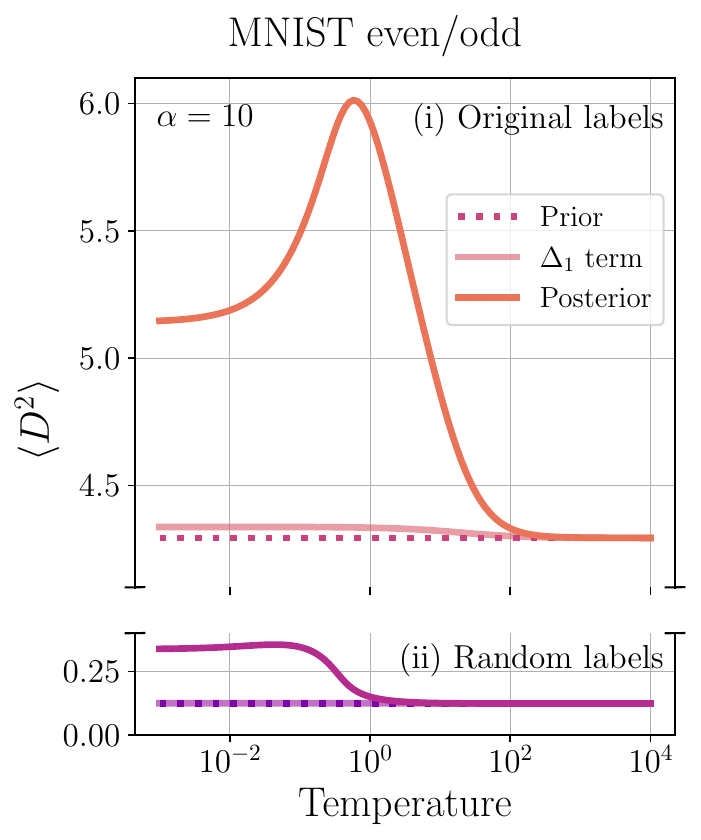}

    \caption{The plots show the inversion peak for the CIFAR-10 and MNIST datasets across different splits. The phenomenology is consistent with that presented in the main text, with an inversion peak that disappears once the labels are randomized. We note that, for the CIFAR-10 dataset split into odd and even classes, the peak struggles to emerge, reflecting the absence of an underlying manifold geometry. All the plots are computed with $P=1000$.}
    \label{fig:six_plots}
\end{figure}

Finally, Fig.~\ref{fig:six_plots} shows that the nonmonotonic behavior of $\langle D^2\rangle$ discussed in the main text is also observed for other class splits of the CIFAR-10 and MNIST datasets, and that it is suppressed when the labels are randomized.

\section{Derivation of the second-order statistics of the weights}\label{sec:weights}
To compute the second order statistics of the internal weights, it is possible to introduce a partition function with a source term:
\begin{equation}
\mathcal{Z}(\{J\}) = \int dvdW e^{-\frac{\beta}{2}|f-y|^2}e^{-\frac{\lambda_0}{2}||W||^2}e^{-\frac{\lambda_1}{2}||v||^2}e^{-\frac{\lambda_0}{2}\left(\sum_{k=1}^{N_1}J_kW_{1k}\right)^2}.
\end{equation}
With this definition, we have
\begin{equation}
    \langle W_{1k}W_{1h}\rangle = -\frac{1}{\lambda_0}\frac{1}{\mathcal{Z}(0)}\frac{\partial^2\mathcal{Z}(\{J\})}{\partial J_k\partial J_h}\Bigg|_{J=0}.
\end{equation}
Note that, being interested in mean values, it is possible not to take care of the normalizations (unless they are $J$-dependent) while computing the partition function.
Considering the inner integral, we separate the contribution of the terms with $i \neq 1$ from that with $i = 1$ (omitting the products over the indices that are being integrated in an obvious way). 
\begin{equation}
    \int \prod_{i\neq1}^{N_1}dW_{ij}\delta\left(h_i^\mu-\frac{1}{N_0}\sum_{j}W_{ij}x_j^\mu\right)
    e^{-\frac{\lambda_0}{2}||W||^2}
    \int dW_{1j} \delta\left(h_1^\mu-\frac{1}{N_0}\sum_{j}W_{1j}x_j^\mu\right)
    e^{-\frac{\lambda_0}{2}||W||^2}
    e^{-\frac{\lambda_0}{2}\left(\sum_{k=1}^{N_1}J_kW_{1k}\right)^2}.
\end{equation}
After the integration, the previous expression assumes the form:
\begin{equation}
    \left( \prod_{i\neq1}^{N_1}
    \frac{1}{\sqrt{\text{Det}(C)}}
    e^{-\frac{1}{2}\sum_{\mu\nu}h_i^\mu C^{-1}_{\mu\nu}h^{\nu}_i} \right)
    \frac{1}{\sqrt{\text{Det}(D(J))}}
    e^{-\frac{1}{2}\sum_{\mu\nu}h^{\mu}_1D(J)^{-1}_{\mu\nu}h^{\nu}_1}
    \frac{1}{\sqrt{\text{Det}(\delta_{ij}+J_iJ_j)}},
\end{equation}
where we defined $D(J)$ as:
\begin{align}
    D(J)_{\mu\nu} & = \sum_{ij}\frac{1}{N_0\lambda_0}x_{i}^\mu x_j^\nu\left(\delta_{ij}+J_iJ_j\right)^{-1} =    \nonumber\\
    & = C_{ij} - \frac{1}{N_0\lambda_0}\frac{1}{1+J^2}\left(\sum_{i}J_ix_i^\mu\right)\left(\sum_{i}J_ix_i^\nu\right),
\end{align}
The inverse is computed by employing the Sherman-Morrison formula. Note that we defined $J^2 = \sum_kJ_k^2$ and that $\text{Det}(\delta_{ij}+J_iJ_j) = 1+J^2$. That said, the partition function is simplified by the introduction of an integral representation of a Dirac delta function $\prod_\mu\delta\left(s^\mu-f(v,h^\mu)\right)$. Integrating the read-out weights, we obtain:
\begin{align} \label{PartitionFunction}
    \mathcal{Z}(\{J\}) &= \int ds d\bar{s}
    e^{-\frac{\beta}{2}|s-y|^2}e^{is\bar{s}}
    \int \prod_{i\neq1}^{N_1}dv_idh_{i}^\mu 
    e^{-\frac{\lambda_1}{2}||v||^2}
    e^{-\frac{1}{2}\sum_{\mu\nu}h_i^\mu C^{-1}_{\mu\nu}h^{\nu}_i}
    e^{-iv_i\frac{1}{\sqrt{N_1}}\sum_{\mu}\bar{s}^\mu\sigma(h^\mu_i)} \nonumber\\
    & \hspace{2cm}
    \int dv_1dh_1^{\mu}e^{-\frac{\lambda_1}{2}v_1^2}
    \frac{1}{\sqrt{\text{Det}(D(J))}}
    e^{-\frac{1}{2}\sum_{\mu\nu}h^{\mu}_1D(J)^{-1}_{\mu\nu}h^{\nu}_1}
    e^{-iv_1\frac{1}{\sqrt{N_1}}\sum_{\mu}\bar{s}^\mu\sigma(h^\mu_1)} 
    \frac{1}{\sqrt{1+J^2}} = \nonumber\\
    & = \int ds d \bar{s}
    e^{-\frac{\beta}{2}|s-y|^2}e^{is\bar{s}}
    \left[1+\frac{1}{\lambda_1N_1}\bar{s}K\bar{s}\right]^{-\frac{N_1-1}{2}}\frac{1}{\sqrt{1+J^2}}
    \left[1+\frac{1}{\lambda_1N_1}\bar{s}K(J)\bar{s}\right]^{-\frac{1}{2}},
\end{align}
where the Kernel $K(J)$ is defined by
\begin{align}
    K(J)_{\mu\nu} &= \int dh\sigma(h^\mu)\sigma(h^\nu)\frac{e^{-\frac{1}{2}hD(J)^{-1}h}}{\sqrt{\text{Det}(D(J))}} = \nonumber\\
    & = \int dhd\bar{h}\sigma(h^\mu)\sigma(h^\nu)e^{-\frac{1}{2}\bar{h}D(J)\bar{h}}e^{ih\bar{h}}.
\end{align}
A straightforward computation of the derivatives yields
\begin{align}
    \partial_h\partial_k &\frac{1}{\sqrt{1+J^2}}\left[1+\frac{1}{\lambda_1N_1}\bar{s}K(J)\bar{s}\right]^{-\frac{1}{2}}\Bigg|_{J=0} = \nonumber\\
    & = -\delta_{hk}\Big[1+\frac{1}{\lambda_1N_1}\bar{s}K\bar{s}\Big]^{-\frac{1}{2}} 
    -\frac{1}{2}\Big[1+\frac{1}{\lambda_1N_1}\bar{s}K\bar{s}\Big]^{-\frac{3}{2}}
    \frac{1}{\lambda_1N_1}\bar{s}\Big(\partial_h\partial_kK(J)\Big|_{J=0}\Big)\bar{s}, 
\end{align}
where the second derivative of the Kernel assumes the form of
\begin{equation}\label{DDKernel}
    \Big(\partial_h\partial_kK(J)\Big|_{J=0}\Big)_{\mu\nu} = 
    \int dhd\bar{h}\sigma(h^\mu)\sigma(h^\nu)
    \left(\bar{h}\frac{x_hx_k^T}{\lambda_0N_0}\bar{h}\right)
    e^{-\frac{1}{2}\bar{h}C\bar{h}}e^{ih\bar{h}}.
\end{equation}
The Fourier variable $\bar{h}$ can be integrated out by means of a translation, leaving
\begin{align}\label{DefinizioneDeltaK}
    \hspace{-1.5cm}\Big(\partial_h\partial_k K(J)\Big|_{J=0}\Big)_{\mu\nu} &=  \frac{1}{\lambda_0N_0}
    \left(\sum_{\lambda\sigma}x_h^\lambda x_k^\sigma C_{\lambda\sigma}^{-1}\right)
    \int dh e^{-\frac{1}{2}hC^{-1}h}\sigma(h^\mu)\sigma(h^\nu) 
    - \nonumber\\
    & \hspace{1.5cm}- \frac{1}{\lambda_0 N_0}\sum_{\eta\delta\lambda\sigma}C^{-1}_{\eta\lambda}x_h^\lambda x_k^\sigma C^{-1}_{\sigma\delta}
    \int dh e^{-\frac{1}{2}hC^{-1}h}\sigma(h^\mu)\sigma(h^\nu)h^\eta h^\delta
\end{align}
Note that the expression appears to be ill-defined since it depends on the single components of the inverse of the covariance matrix, which is non-existent as soon as $P>N_0$, which is typically the case of interest. It turns out that it is not the case and that the expression can be recast in such a way that the dependence on the single components of $C^{-1}$ is removed. Note also that the presence of $C^{-1}$ in the Gaussian measure associated with the variable $h$ does not represent an issue since the distribution is well defined through its Fourier transform.
To show the independence of the previous expression on $C^{-1}$, we first note that the integrals define two different kernels: while the first is the usual NNGP kernel, which we denote $K_{\mu\nu}$, the second depends on more indices and is defined as
\begin{equation}
    K_{\mu\nu}^{\eta\delta} = \int dh e^{-\frac{1}{2}hC^{-1}h}\sigma(h^\mu)\sigma(h^\nu)h^\eta h^\delta .
\end{equation}
To recast this expression, we express it in terms of the NNGP kernel. To do so, we introduce a source such that:
\begin{align}
    {K}^{\eta\delta}_{\mu\nu} &= \partial_\eta\partial_\delta(K(\{J\}))^{\mu\nu}\Big|_{J=0} =  \nonumber \\
    & =\partial_\eta\partial_\delta \int dh e^{-\frac{1}{2}hC^{-1}h}\sigma(h^\mu)\sigma(h^\nu)e^{\sum_\alpha J_\alpha h^\alpha}\Bigg|_{J=0}.
\end{align}
After completing the square and performing a translation, the kernel is brought in the form of:
\begin{align}
    (K(\{J\}))^{\mu\nu} 
    = e^{\frac{1}{2}JCJ}\int dh e^{-\frac{1}{2}hC^{-1}h}\sigma\left(h^\mu-(CJ)^\mu\right)\sigma\left(h^\nu-(CJ)^\nu\right) .
\end{align}
The computation of the derivatives returns:
\begin{align}
    K^{\lambda\sigma}_{\mu\nu} &=  C_{\lambda\sigma} K _{\mu\nu} + C_{\mu\sigma}C_{\mu\lambda}\int \mathcal{D}h \hspace{2pt}\sigma''_\mu\sigma_\nu + C_{\nu\lambda}C_{\nu\sigma}\int \mathcal{D}h \hspace{2pt}\sigma''_\nu\sigma_\mu + \left(C_{\mu\lambda}C_{\nu\sigma}+ C_{\nu\lambda}C_{\mu\sigma}\right)\int \mathcal{D}h \hspace{2pt} \sigma'_\mu\sigma'_\nu,
\end{align}
where $\sigma'_\mu$ and $\sigma''_\mu$ are the first and second derivatives of the activation function evaluated on $h^\mu$.
Inserting this expression in Eq.~\eqref{DefinizioneDeltaK}, the sums over the Greek indices remove the dependence on $C^{-1}$:
\begin{align}
    \Big(\partial_h\partial_k K(J)\Big|_{J=0}\Big)_{\mu\nu}
    &= -\int dh e^{-\frac{1}{2}hC^{-1}h}\left[\sigma''(h^\mu)\sigma(h^\nu)x^\mu_hx^\mu_k + \sigma(h^\mu)\sigma''(h^\nu)x^\nu_hx^\nu_k\right] - \nonumber \\
    &\hspace{1.5cm}-\frac{x^\mu_h x^\nu_k+x^\nu_hx^\mu_k}{\lambda_0N_0}\int dh e^{-\frac{1}{2}hC^{-1}h}\sigma'(h^\mu)\sigma'(h^\nu),
\end{align}
which is well defined also in case of $P>N_0$. Defining 
\begin{equation}
    (K'')_{hk}^{\mu\nu} = \int dh e^{-\frac{1}{2}hC^{-1}h}\left[\sigma''(h^\mu)\sigma(h^\nu)x^\mu_hx^\mu_k + \sigma(h^\mu)\sigma''(h^\nu)x^\nu_hx^\nu_k\right],
\end{equation}
\begin{equation}
    (K')^{\mu\nu} = \int dh e^{-\frac{1}{2}hC^{-1}h}\sigma'(h^\mu)\sigma'(h^\nu),
\end{equation}
we denote for convenience $\Big(\partial_h\partial_k K(J)\Big|_{J=0}\Big)_{\mu\nu} = \Delta_{hk}^{\mu\nu}$, with:
\begin{equation}
    \Delta_{hk}^{\mu\nu} = - (K'')^{\mu\nu}_{hk}-\frac{x^\mu_h x^\nu_k+x^\nu_hx^\mu_k}{\lambda_0N_0}(K')^{\mu\nu}.
\end{equation}
It is now possible to continue with the computation of the second order statistics. From the definition and the previous results, we have:
\begin{align}
    \langle W_{1h}W_{1k}\rangle &= \frac{\delta_{hk}}{\lambda_0\mathcal{Z}(0)}
    \int ds d \bar{s}
    e^{-\frac{\beta}{2}|s-y|^2}e^{is\bar{s}}
    \left[1+\frac{1}{\lambda_1N_1}\bar{s}K\bar{s}\right]^{-\frac{N_1}{2}} + \nonumber\\
    & \hspace{1.5cm}+ \frac{1}{2\lambda_0\mathcal{Z}(0)}
    \int ds d \bar{s}
    e^{-\frac{\beta}{2}|s-y|^2}e^{is\bar{s}}
    \left[1+\frac{1}{\lambda_1N_1}\bar{s}K\bar{s}\right]^{-\frac{N_1}{2}-1}\frac{1}{\lambda_1N_1}\bar{s}\Delta_{hk}\bar{s}.
\end{align}
From Eq.~\eqref{PartitionFunction}, by plugging $J=0$ one can check that the integral in the first term is exactly $\mathcal{Z}(0)$, so that the first contribution to the expectation value is simply $\delta_{hk}/\lambda_0$. To make further progress with the second term, we use the integral definition of the Gamma function (the Gamma itself plays the role of a normalization constant and is not reported):
\begin{align}
    \left[1+\frac{\bar{s}K\bar{s}}{\lambda_1N_1}\right]^{-\frac{N_1}{2}-1} &= \int_{\tilde{Q}>0}d\tilde{Q}e^{-\tilde{Q}\left(1+\frac{\bar{s}K\bar{s}}{\lambda_1N_1}\right)}\tilde{Q}^{\frac{N_1}{2}} = \nonumber\\
    &=\int_{Q>0}dQe^{-\frac{N_1}{2}Q\left(1+\frac{\bar{s}K\bar{s}}{\lambda_1N_1}\right)}Q^{\frac{N_1}{2}-1}Q
\end{align}
The second line is obtained by rescaling the integration variable $\tilde{Q}=QN_1/2$. The integral in the second term reads:
\begin{align}
    & \int_{Q>0} dQdsd\bar{s} e^{-\frac{N_1}{2}Q\left(1+\frac{\bar{s}K\bar{s}}{\lambda_1N_1}\right)}Q^{\frac{N_1}{2}-1}Q
    e^{-\frac{\beta}{2}|s-y|^2}e^{is\bar{s}}
    \frac{\bar{s}\Delta_{hk}\bar{s}}{\lambda_1N_1} =\nonumber\\
    & = \int_{Q>0} dQ e^{-\frac{N_1}{2}Q-\frac{N_1-2}{2}\log Q}\int d\bar{s}e^{-\frac{1}{2}(\bar{s}+i\tilde{K}_{(R)}^{-1}y)\tilde{K}_{(R)}(\bar{s}+i\tilde{K}_{(R)}^{-1}y)}
    e^{-\frac{1}{2}\log\beta}
    e^{-\frac{1}{2}y\tilde{K}_{(R)}^{-1}y}\frac{Q}{\lambda_1N_1}\bar{s}\Delta_{hk}\bar{s} = \nonumber\\
    & = \int_{Q>0} dQ e^{-\frac{N_1}{2}Q-\frac{N_1-2}{2}\log Q}e^{-\frac{1}{2}\log\beta}e^{-\frac{1}{2}y\tilde{K}_{(R)}^{-1}y}
    \int d\bar{t}e^{-\frac{1}{2}\bar{t}\tilde{K}_{(R)}\bar{t}}\frac{Q}{\lambda_1N_1}\left(
    \bar{t}-i\tilde{K}_{(R)}y
    \right)
    \Delta_{hk}
    \left(
    \bar{t}-i\tilde{K}_{(R)}y
    \right) = \nonumber\\
    & = \int_{Q>0} dQ e^{-\frac{N_1}{2}Q-\frac{N_1-2}{2}\log Q}e^{-\frac{1}{2}y\tilde{K}_{(R)}^{-1}y}e^{-\frac{1}{2}\log\text{Det}\beta\tilde{K}_{(R)}}
    \frac{Q}{\lambda_1N_1}\sum_{\mu\nu}\left[
    (\tilde{K}^{-1}_{(R)})_{\mu\nu}\Delta_{hk}^{\mu\nu}-\sum_{\lambda\sigma}(\tilde{K}^{-1}_{(R)})_{\mu\lambda}y_\lambda(\Delta_{hk}^{\mu\nu})
    (\tilde{K}^{-1}_{(R)})_{\nu\sigma}y_\sigma
    \right] = \nonumber \\
    & = \int_{Q>0} dQ e^{-\frac{N_1}{2}S(Q)} \frac{Q}{\lambda_1N_1}\sum_{\mu\nu}\left[
    (\tilde{K}^{-1}_{(R)})_{\mu\nu}\Delta_{hk}^{\mu\nu}-\sum_{\lambda\sigma}(\tilde{K}^{-1}_{(R)})_{\mu\lambda}y_\lambda(\Delta_{hk}^{\mu\nu})
    (\tilde{K}^{-1}_{(R)})_{\nu\sigma}y_\sigma
    \right],
\end{align}
where the effective action $S(Q)$ is defined as
\begin{equation}
        S(Q) = -Q+\frac{N_1-2}{N_1}\log{Q}-\frac{\alpha}{P}y^T\tilde{K}_{(R)}^{-1}y-\frac{\alpha}{P}\text{Tr}\log \beta\tilde{K}_{(R)},
\end{equation}
with 
\begin{equation}
    \tilde{K}_{(R)} = \frac{\mathds{1}}{\beta} + \frac{Q}{\lambda_1}K.
\end{equation}
By introducing an integration variable $Q$ in the same way as was done in the previous lines, one can easily check that the partition function $\mathcal{Z}(0) = \int_{Q>0}\text{exp}[-N_1S(Q)/2]$. Because of that, in the proportional limit where $N_1$ is large, the integral is dominated by the saddle-point contribution ($\bar{Q}$ such that $\partial_QS(Q)|_{Q=\bar{Q}} = 0$), returning:
\begin{equation}
    \langle W_{1h}W_{1k}\rangle = \frac{\delta_{hk}}{\lambda_0} + \frac{1}{\lambda_0}\frac{\bar{Q}}{2\lambda_1N_1}
    \sum_{\mu,\nu=1}^{P}\left[
    (\tilde{K}^{-1}_{(R)})_{\mu\nu}\Delta_{hk}^{\mu\nu}-\sum_{\lambda\sigma}(\tilde{K}^{-1}_{(R)})_{\mu\lambda}y_\lambda(\Delta_{hk}^{\mu\nu})
    (\tilde{K}^{-1}_{(R)})_{\nu\sigma}y_\sigma
    \right].
\end{equation}
As mentioned in the main text, the first term is obtained as the expectation value over the prior distribution of the weights $W$: $\langle W_{1h}W_{1k}\rangle_{P(W)}=\delta_{hk}/\lambda_0$ or over the posterior in the infinite-width limit. The latter statement can be easily checked by noting that when $P$ is finite the term proportional to $1/N_1\sum^P\rightarrow0$. In the proportional limit, as argued in the main text, this term can be nontrivial and bring finite corrections to the two-point function.

\section{Consistency check for linear activation function}\label{sec:linear}
In the case of a linear activation function $\sigma = Id$, we can compute
\begin{equation}\label{SimilarityLinear}
    \langle\sigma(h^\mu_1)\sigma(h^\mu_1)\rangle = \langle h_1^\mu h_1^\nu\rangle = \frac{1}{N_0}\sum_{hk}x_h^\mu x_k^\nu\langle W_{1h}W_{1k}\rangle 
\end{equation}
and the result must match the one presented in Eq.~\eqref{SimilarityMatrix}, which is obtained by an independent computation. First of all, the expression of $\Delta_{hk}^{\mu\nu}$ simplifies as long as $\sigma'(h^\mu)=1$ and $\sigma''(h^\mu)=0$, returning
\begin{equation}
    \Delta_{hk}^{\mu\nu} = -\frac{x_h^\mu x_k^\nu + x_h^\nu x_k^\mu}{\lambda_0 N_0}.
\end{equation}
Furthermore, $K_{\mu\nu} = C_{\mu\nu}$. Plugging the previous equation into the expectation value:
\begin{equation}
    \langle W_{1h}W_{1k}\rangle = \frac{\delta_{hk}}{\lambda_0}+\frac{1}{\lambda_0}\frac{\bar{Q}}{\lambda_1N_1}\sum_{\lambda\sigma}\left[-(\tilde{C}^{-1}_{(R)})_{\lambda\sigma}\frac{x_h^\lambda x_k^\sigma}{\lambda_0N_0} + (\tilde{C}_{(R)}^{-1}y)_\lambda(\tilde{C}_{(R)}^{-1}y)_\sigma \frac{x_h^\lambda x_k^\sigma}{\lambda_0N_0}
    \right],
\end{equation}
where we used the notation $\tilde{C}_{(R)} = \mathds{1}/\beta + \bar{Q}C/\lambda_1$. The computation of $\langle h_i^\mu h_i^\nu\rangle$ involves two further sums over the indices $h,k$. The terms involved return:
\begin{equation}
    \sum_{hk}\frac{x_h^\mu x_k^\nu}{N_0}\frac{\delta_{hk}}{\lambda_0} = C_{\mu\nu},
\end{equation}
\begin{equation}
    \frac{1}{\lambda_0}\sum_{hk}-\frac{x_h^\lambda x_k^\sigma}{\lambda_0 N_0}\frac{x_h^\mu x_k^\nu}{N_0} = -C_{\mu\lambda}C_{\nu\sigma}.
\end{equation}
With this in mind, plugging the previous expressions in Eq.~\eqref{SimilarityLinear}, we obtain:
\begin{equation}
    \langle h_1^\mu h_1^\nu\rangle = C_{\mu\nu} -\frac{\bar{Q}}{\lambda_1N_1}\sum_{\lambda\sigma}^P C_{\mu\lambda}C_{\nu\sigma}\left[(\tilde{C}^{-1}_{(R)})_{\lambda\sigma}-(\tilde{C}_{(R)}^{-1}y)_\lambda(\tilde{C}_{(R)}^{-1}y)_\sigma \right],
\end{equation}
which is the equivalent of Eq.~\eqref{SimilarityMatrix} in the case of linear activation function.

\end{document}